\documentclass[aps,prl,twocolumn,superscriptaddress,showpacs]{revtex4-1}
\usepackage{bibunits}
\usepackage{subfigure}
\usepackage{times}
\usepackage{amsfonts}
\usepackage{amssymb}
\usepackage{amsmath}
\usepackage{graphicx}
\usepackage{color}

\begin{document}
\title{Reversible first-order transition in Pauli percolation}

\author{Mykola Maksymenko}

\affiliation{Max-Planck-Institut f\"{u}r Physik komplexer Systeme,
        N\"{o}thnitzer Stra{\ss}e 38, 01187 Dresden, Germany}
\author{Roderich Moessner}
\affiliation{Max-Planck-Institut f\"{u}r Physik komplexer Systeme,
        N\"{o}thnitzer Stra{\ss}e 38, 01187 Dresden, Germany}
        \author{Kirill Shtengel}
\affiliation{Max-Planck-Institut f\"{u}r Physik komplexer Systeme,
        N\"{o}thnitzer Stra{\ss}e 38, 01187 Dresden, Germany}
\affiliation{Department of Physics and Astronomy, University of California at
Riverside, Riverside,  CA 92521, USA}
\date{\today}



\begin{abstract}
 Percolation plays an important role in fields and phenomena as diverse as
the study of social networks, the dynamics of epidemics, the robustness of
electricity grids, conduction in disordered media, and geometric properties
in statistical physics. We analyse a new percolation problem in which the
\emph{first order} nature of an \emph{equilibrium} percolation transition can
be established analytically and verified numerically. The rules for this site
percolation model are physical and very simple, requiring only the
introduction of a weight $W(n)=n+1$ for a cluster of size $n$.  This
establishes that a discontinuous percolation transition can occur with
qualitatively more local interactions than in all currently considered
examples of explosive percolation; and that, unlike these, it can be
reversible. This greatly extends both the applicability of such percolation
models in principle, and their reach in practice.
\end{abstract}

\maketitle
\defaultbibliographystyle{Science}
\defaultbibliography{kirref_science}
\begin{bibunit}[Science]
{\emph{Introduction. }}The percolation transition involves fundamentally
geometric properties, manifest in non-local observables such as an onset of
conductivity in a dirty metal, a breakdown of an electrical grid or an epidemic
disease outbreak
\cite{Isichenko1992,Stauffer1994,Moore200epidemics,Golnoosh2012}. This is at
odds with the more standard phase transitions in statistical physics which are
described by a local order parameter, such as the magnetisation in a bar
magnet. It thus involves a conceptually fundamentally distinct set of issues.
Its wide applicability coupled with this fundamental importance have generated
much interest in defining various types of percolation problems and analysing
their concomitant phase transitions. One enterprise has been the quest for a
first-order percolation transition, where the percolating cluster sets in
discontinuously, corresponding to a particularly violent transition, which can
qualitatively amplify desirable properties in applications. Such a transition
has remained remarkably elusive, but the development which has taken
place under the heading of \emph{explosive percolation} has finally yielded
one, via a mechanism in which an infinite number of nonlocal interactions need
to occur simultaneously
\cite{Achlioptas2009,daCosta2010,Riordan2011,Ziff2009,Pan2011,Grassberger2011,
Panagiotou2013,Manna2010,Cho2013}.

Here, we study \emph{Pauli percolation} -- a site percolation problem with its
origin in correlated quantum magnetism, characterized by a number of novel
striking and desirable properties. First of all, it exhibits a first-order
phase transition invoking only a minimal amount of non-locality, in the form of
an interaction solely between adjacent clusters, depending only on their
respective sizes. Secondly, such an interaction can be very easily generated
from perfectly local ones, for instance either via a simple classical colouring
rule, or via a quantum-statistical interaction between Fermionic particles.
Thirdly, it describes an equilibrium phase transition, and is hence reversible;
at the same time, it can be thought of and analysed as a stochastic dynamical
process and thus may -- but need not -- exhibit hysteresis. Finally, Pauli
percolation lends itself to investigations using the toolbox of equilibrium
classical statistical mechanics; we are thus able to solve its properties
\emph{analytically} on a regular random graph, and verify this solution via
numerical Monte Carlo simulations.

{\emph{Pauli percolation. }}
\begin{figure}
\subfigure[]{\includegraphics[width=0.6\columnwidth]{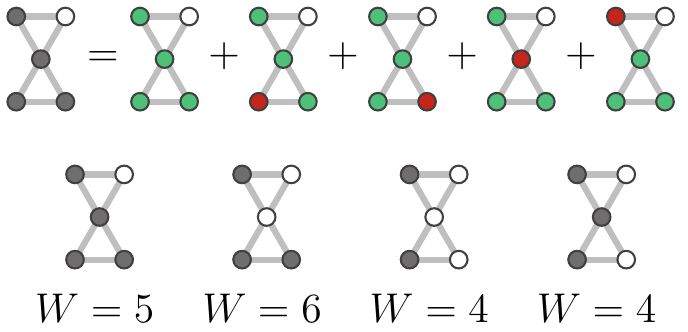}
\label{fig:weights}
}
\subfigure[]{\includegraphics[width=0.9\columnwidth]{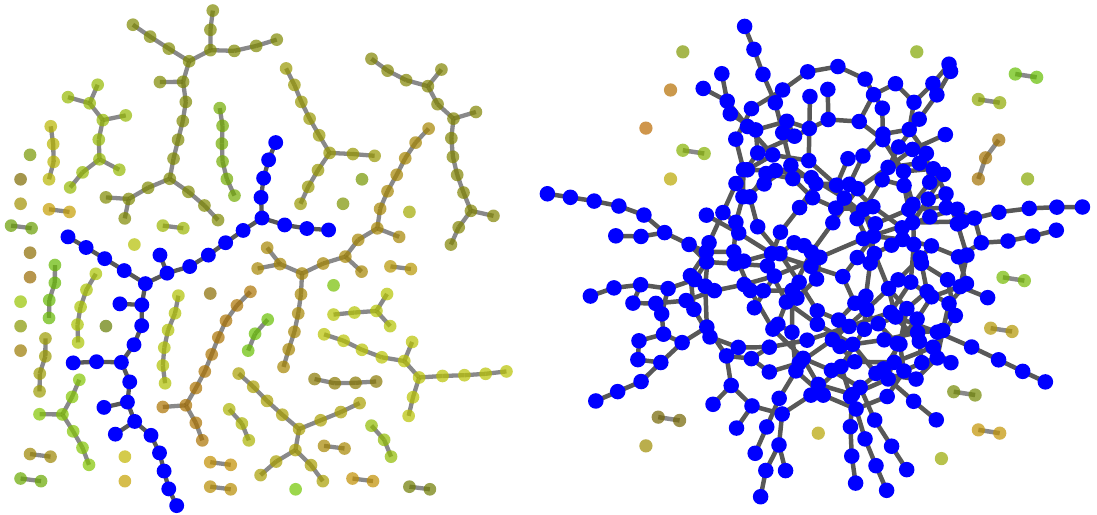}
\label{fig:first_order_percolation}
}
\caption
{(a) In Pauli percolation, weight ${W=n+1}$ of a cluster can be
reproduced by imposing a simple two-color `contagion' rule shown here: the
whole cluster of occupied sites can be either healthy (green) or have a single
infected site (red).
Different cluster configurations appear with different statistical weights. (b)
The explosive nature of a Pauli-percolation transition on a regular random graph
of $N=400$ sites: two representative configurations, without and with a giant cluster
at the same site fugacity corresponding to $\tilde{p}=0.45$ are shown side by side.
The largest cluster is colored blue; unoccupied sites are not shown.
}
\label{fig:process_lattice}
\end{figure}
The model we consider first arose in a quantum many-body problem of itinerant
electrons on lattices with flat energy bands. Such a system can exhibit
flat-band ferromagnetism: the Pauli exclusion principle mandates that in the
ground state the electron spins in a cluster order ferromagnetically in order
to minimize the energy of repulsive on-site interactions~\cite{Mielke1993}.
This leads to a weight of $(n+1)$, reflecting the number of possible
orientations of the total spin of a ferromagnetic cluster of $n$ electrons
\cite{Maksymenko2012}.

The corresponding statistical-mechanical problem describes $M$ particles
occupying random sites of a lattice. Every configuration
$\mathcal{C}=\cup_i \mathcal{C}_i$ appears with statistical weight
$W_{\mathcal{C}}=\prod_i(n(\mathcal{C}_{i})+1)$, with $n(\mathcal{C}_{i})$
being the size of cluster $\mathcal{C}_{i}$. The partition function is
therefore $Z = \sum_{\{\mathcal{C}\}}W_{\mathcal{C}}.$

Merging two clusters of size $m$ and $n$ reduces their overall weight from
$(m+1)(n+1)$ to $(m+n+1)$ -- a dramatic reduction for large clusters resulting
in an effective repulsive interaction between them. This is reminiscent of the
`product rule' leading to explosive percolation suggested by Achlioptas
\cite{Achlioptas2009} and developed in \cite{Manna2010, Cho2013, Ziff2013} but
there are fundamental differences, see discussion below.

Rather then fixing the number of occupied sites, we can study the grand
canonical ensemble by letting each site of the lattice be occupied
with an \emph{a priori} probability ${p}$ or left empty with an a priori
probability $1-{p}$. The grand canonical partition function is then
\begin{equation}
Z = \sum_{\{\mathcal{C}\}}\left(\frac{p}{1-p}\right)^{n(\mathcal{C})}
W_{\cal C}
\label{eq:grandpartition1}
\end{equation}
where $\ln{\left[{p}/(1-{p})\right]}$ plays the role of a chemical potential
controlling the density of occupied sites and letting it fluctuate. Note that
a~priori probability ${p}$, unlike a regular site percolation, is not equal
to the density of occupied sites.

This model also has a simple representation as a particular classical
two-color, or contagion, percolation problem. It is a mild variation of regular
percolation: sites can come in two colors, green (uninfected) or red
(infected). Specifically, each site of a lattice is occupied and colored either
green or red with an a priori probability $\tilde{p}$ each, or left empty with
an a priori probability $1-2\tilde{p}$. Only configurations $\{{\mathcal{G}}\}$
where every cluster contains no more than one red site are taken into account.
The partition function of this model is then simply
\begin{equation}
Z_{\text{gr}}\!=\!\sum_{{\{{\mathcal{G}}}\}}
\left(\frac{\tilde{p}}{1-2\tilde{p}}\right)^{n({\mathcal{G}})}
\label{eq:grandpartition2}
\end{equation}
It is straightforward to see that tracing over all possible site colors
consistent with fixed site occupations renders Eq.~(\ref{eq:grandpartition2})
identical to Eq.~(\ref{eq:grandpartition1}) (with the identification of
$\tilde{p} = p/(p + 1)$): each cluster may have either all sites green
(uninfected), or \emph{at most} one red (infected) site. Therefore a cluster of
$n$ sites has weight ${(n + 1)}$ after the sum over possible locations of red
sites is taken into account. The utility of the formulation as a two-color
percolation problem lies in the fact that the need ever to compute cluster
sizes is obviated: the choice of location of the infected site takes care of
that.

\begin{figure}
\begin{center}
\subfigure[]{\includegraphics[width=0.9\columnwidth]{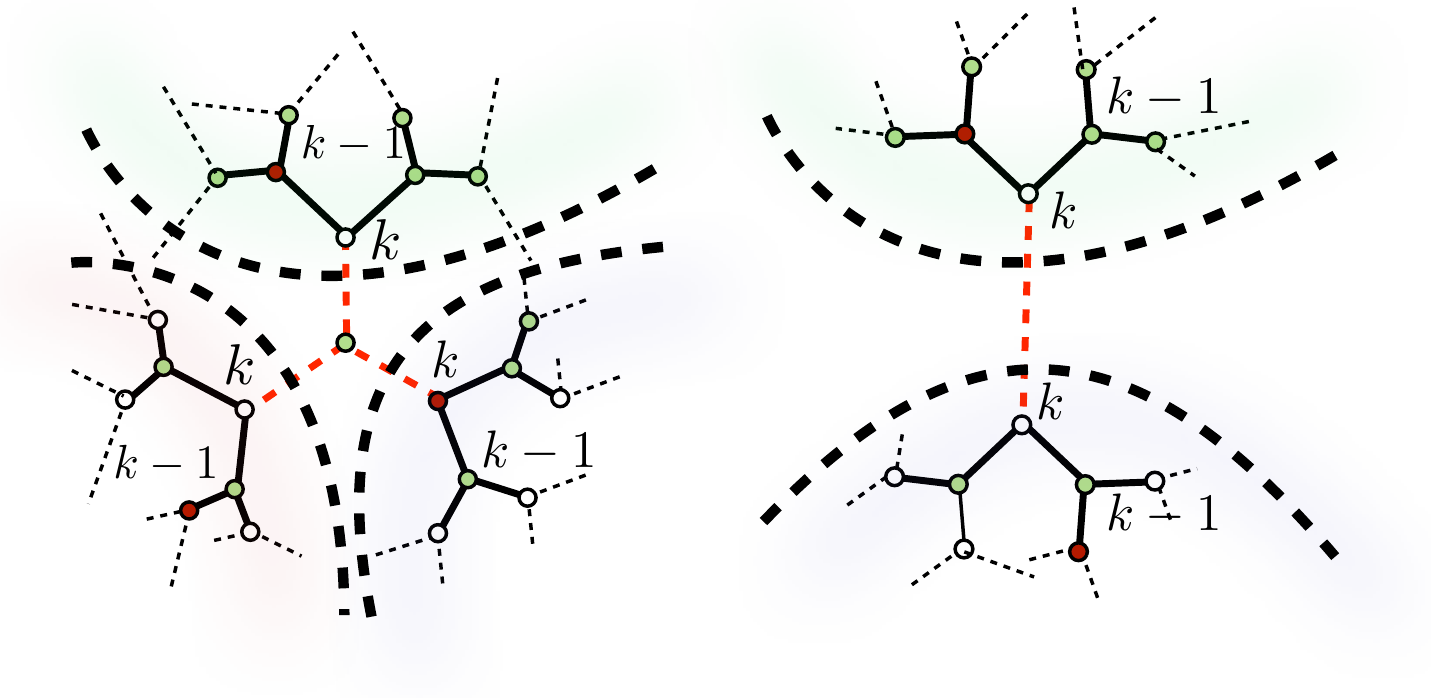}
\label{fig:bethe}}
\subfigure[]{\includegraphics[width=0.9\columnwidth]{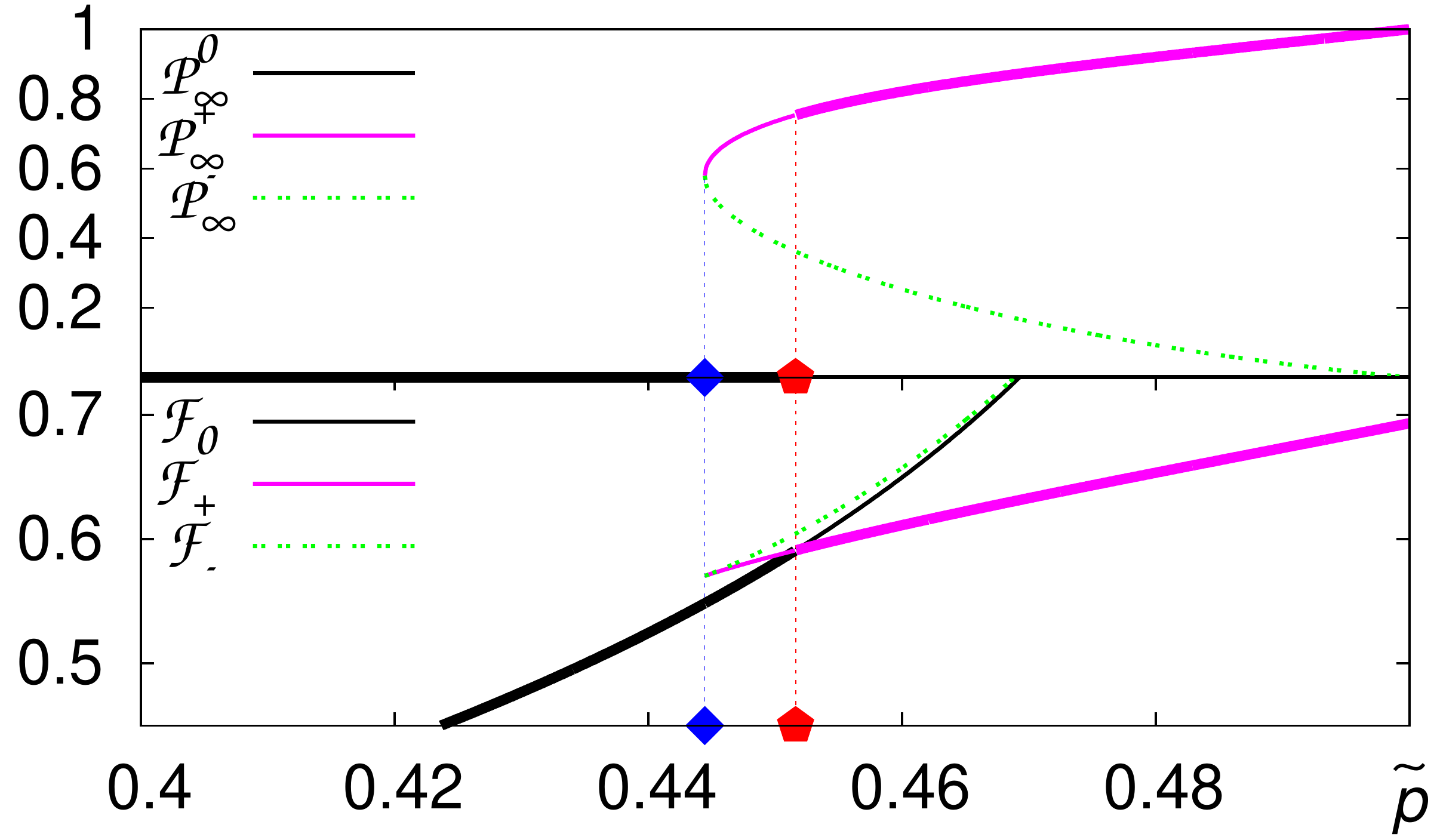}
\label{fig:percolation}}
\caption
{(a) Properties of a regular random graph make it locally equivalent to the
neighborhood of an internal site of a Bethe lattice obtained by
connecting roots of independent Cayley trees via site or edge
addition. Here the sites colors (red/dark grey and green/light grey) represent
one instance allowed by the two-color contagion percolation rules.
(b) The upper pane shows the probability ${\mathcal{P}}_{\infty}$ that a site belongs
to a giant cluster as a function of
$\tilde{p}$. The blue diamond marks $\tilde{p}=4/9$ at which
the nonzero solution appears.
 The lower pane shows the `bulk' free energy per site of a 3-regular random graph
 corresponding to each of these solutions (see text for details).
The red pentagon indicates the transition point. Bold parts of the lines in both
panels indicate the
actual solution.
}
\label{fig02}
\end{center}
\end{figure}

{\emph{Analytic and numerical results. }} We show that Pauli percolation
exhibits a discontinuous percolation transition in infinite dimensions by
studying it analytically and numerically on a regular random graph of ${N}$
sites. Such graphs are often used to approximate random networks
\cite{Dorogovtsev2010book}. They have a vanishing density of short cycles and
mostly contain loops of size $\ln{N}$; hence they are locally tree-like
\cite{Mezard2001, Janson2000}. This property enables us to develop an exact
solution via a so-called \emph{cavity method} widely used in spin glass and
optimization problems \cite{Mezard2001, Laumann2008, Rivoire2005,Barre2007,
Hsu2013}. In the cavity method, adding a site or an edge to a $z$-regular
random graph is equivalent to connecting $z$ or $z-1$ roots (here referred to
as cavity sites) of independent Cayley trees (see Figure~\ref{fig:bethe}) via
that site or edge. To complete the correspondence and get the correct set of
solutions we introduce `wired' boundary conditions which connect the outer
sites (`leaves') with one another thus allowing the formation of loops.

The recursive structure of calculations on Cayley trees makes the mean-field
treatment exact in these systems. Care must be taken to correctly calculate the
bulk thermodynamic potentials on such structures \cite{Hoory2006,
Laumann2009,Barre2007,Barre2001}. For instance, the bulk free energy is
computed as a change in free energy due to the addition of a site and the
corresponding links emanating from this site ${\mathcal{F}  =
\lim_{k\to\infty}\left[
-\ln{{\mathcal{Z}}/Z_k^3}+\left(z/2\right)\ln{{\tilde{\mathcal{Z}}}/Z_k^2}\right]}$,
where ${\mathcal{Z}}$ and $\tilde{\mathcal{Z}}$ are the partition functions for
a uniform Bethe lattice obtained by connecting either $z$ or $z-1$ root sites
of independent trees via a new site or edge. $Z_k$ is the partition function
for a level-$k$ tree~\cite{Mezard2001, Laumann2008, Rivoire2005,Barre2007,
Hsu2013}.

In the first instance we are interested in the existence of a giant cluster
(i.e. a cluster occupying a finite fraction of the lattice), in the simplest
case of $z=3$. We define ${\mathcal{P}}_{\infty}$ to be the probability that
a given site belongs to such giant cluster; see Supplementary Material for
details. For $\tilde{p}<4/9$, the only solution of the resulting equations is
${\mathcal{P}}^\text{0}_{\infty}=0$ -- in other words, there is no
percolation. For $\tilde{p}\geq4/9$ two more solutions appear with
${\mathcal{P}}^{\pm}_{\infty}\neq 0$ as shown in Figure~\ref{fig:percolation}
(with the lower brunch being unphysical) . Note that there is never a
percolating uninfected cluster: the probability that a given cluster of size
$n$ remains uninfected is $1/(n+1)$.

The topology of the plot for ${\mathcal{P}}_{\infty}$ already demonstrates the
first order nature of the percolation transition: the curve which yields the
solution $\mathcal{P}^{+}_{\infty} = 1$ (i.e. all sites are occupied) for
$\tilde{p} \to 1/2$
 never crosses the non-percolating solution $\mathcal{P}^\text{0}_{\infty} = 0$,
which in turn is unique for $\tilde{p} \to 0$. The transition from one to the
other therefore implies a jump in $\mathcal{P}_{\infty}$! To determine when the
actual transition takes place we analyze the bulk free energy of the problem.
The solution which minimizes this quantity maximizes the partition function and
thus is selected. This selects the solution of ${\mathcal{P}}^{+}_{\infty}\neq
0$ at $\tilde{p}_{c}=0.451606...$(See Figure~\ref{fig:percolation}) indicating a
discontinuous jump as soon as $\tilde{p}= \tilde{p}_{c}$. We note that this is
in agreement with other quantities such as cluster size distribution and
average cluster size which show no signature of power-law distribution or
divergence at the transition point. These, together with details of the
computation, are shown in the Supplemental Material.

\begin{figure}
\centering
\includegraphics[width=.98\columnwidth]{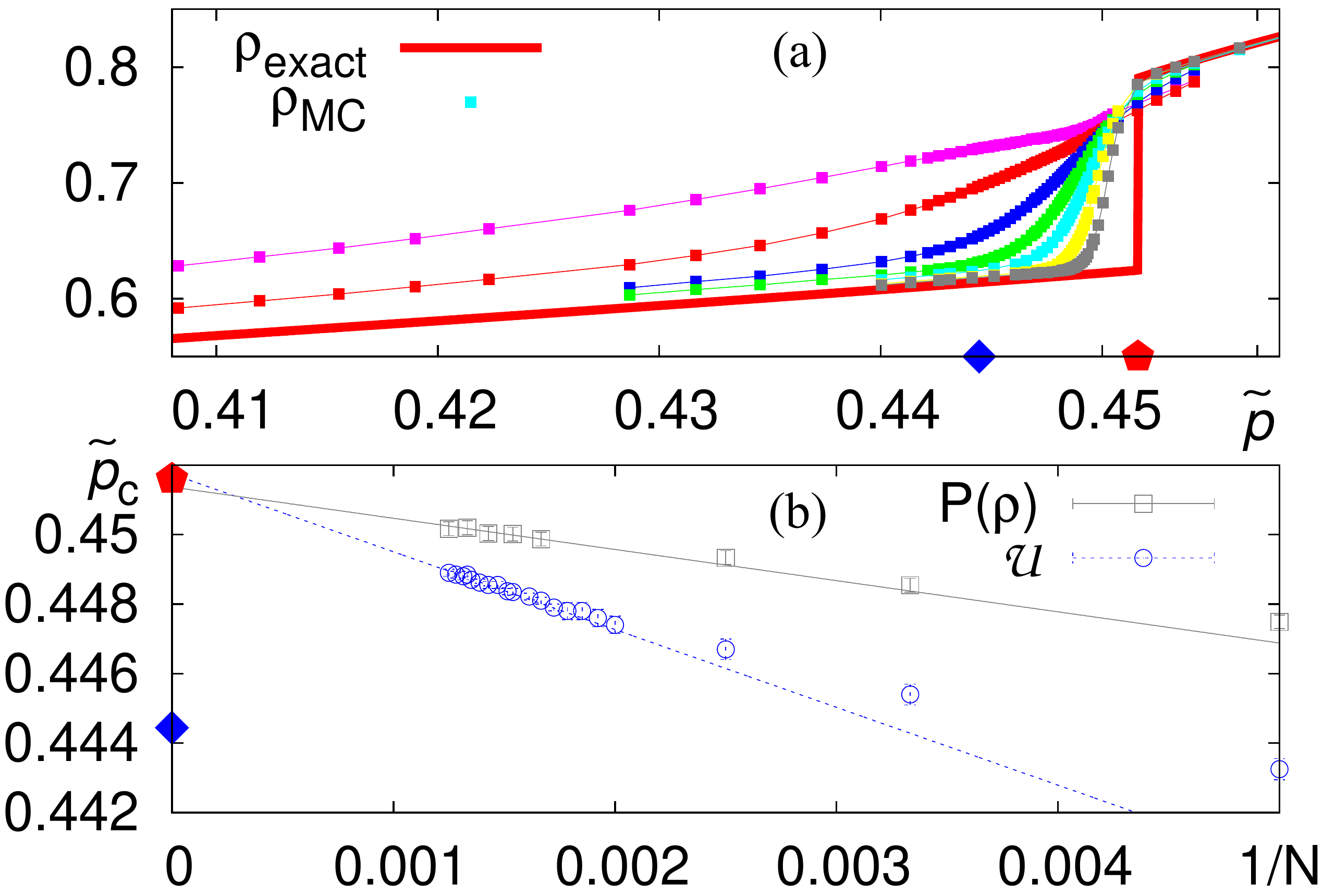}
\caption
{(a) Density $\rho$ versus a priori probability $\tilde{p}$. Red line indicates
the exact solution while the dots represent Monte-Carlo results for 3-regular random graphs
of sizes ${N}=50,100, 200, 300, 400, 600, 800$.
(b) Finite size extrapolation of the minimum of the fourth-order Binder cumulant of
density $\cal{U}$ and point of equally weighed peaks of histograms of density $P(\rho)$. The red pentagon marks the point at which the
infinite cluster appears in the thermodynamic limit, at a density distinct
from where the solutions ${\cal P}^{\pm}_{\infty}$ first appear, indicated by the blue diamond.}
\label{fig03}
\end{figure}

We support our analytic results by Monte-Carlo simulations of
Eq.~(\ref{eq:grandpartition1}) on a regular random graph. We analyze density of
occupied sites $\rho$ as well as histograms of its distribution along with the
fourth-order Binder cumulant $\mathcal{U}$ as standard indicators of phase
transitions. In all the quantities the extrapolation to ${N}\rightarrow\infty$
is consistent with the exact solution. Below and above the transition the
numerical data follows the branches of the exact solution for the uniform Bethe
lattice. The histograms of the density distribution give a clear double peak
structure -- the hallmark of a discontinuous transition -- and in
Figure~\ref{fig03} we provide an extrapolation of the point at which these two
peaks are of equal weight. This nicely extrapolates to the analytic result for
the transition point $\tilde{p}_{c}=0.452(3)$. Finally the density Binder
cumulant $\mathcal{U}$  develops a minimum at the transition point -- a typical
behavior for a discontinuous transition; its extrapolation to thermodynamic
limit is also in good agreement with the transition point obtained
analytically.

{\emph{Discussion. }} The attractiveness of Pauli percolation is manifold.
Firstly, it is underpinned by a simple and transparent physical mechanisms.
Secondly, it is amenable to detailed numerical and analytical analyses.
Thirdly, and crucially, it exhibits a remarkable phenomenology featuring a
\emph{reversible first-order} percolation transition. In the following, we
discuss the import of these items, and embed them in a broader zoology of
percolation problems.

The notion of explosive first-order percolation~\cite{Achlioptas2009} has been
met with much excitement, yet the initial approach proved to be
deficient~\cite{Riordan2011}. A discontinuous transition has finally been found
in several variants of explosive percolation models which, however, require a
very elevated degree of non-locality: a dynamical process defining these models
involves a comparison between an \emph{extensive} number of degrees of freedom
before a configuration change occurs
\cite{daCosta2010,Riordan2011,Grassberger2011,Manna2010}. Recent studies also
considered suppressing the onset of percolation through a rule explicitly
inhibiting bond addition if it leads to the formation of a spanning cluster
\cite{Cho2013}. Not only such a process involves an extensive number of local
degrees of freedom, it also makes the `microscopic' dynamics of the model -- a
placement of a particular bond -- explicitly depend on the onset of a global
phenomenon, percolation.

Pauli percolation, by contrast, considers one site at a time, with a minimal
amount of non-locality entering only via the sizes of the clusters impinging on
the site in question. In other words, Pauli percolation is non-local only up to
the size of the clusters present locally. The Pauli principle of quantum
mechanics presents a straightforward physical origin for such a weight: quantum
statistical interactions are intrinsically non-local on this level. A classical
route to the same weights involves permitting at most one site of each cluster
to be infected, again a simple and intuitive description involving clusters
only locally and individually.

Nor does Pauli percolation require the irreversibility of explosive
percolation. Based on statistical weights of configurations rather than rules
for cluster growth, Pauli percolation provides an \emph{equilibrium}
first-order transition. It in particular allows for shrinking, as well as
growing, clusters. It therefore naturally accommodates healing/repairing
processes, in e.g. network applications which, notably, can remove percolation
discontinuously. The growth process encoded by the ``product rule'' in
explosive percolation is reminiscent of the weights of Pauli percolation: the
latter, however, provides a natural prescription for removing particles as
well. We should note that another route to a reversible first-order percolation
transition, although not normally thought of in these terms, is provided by the
Fortuin--Kasteleyn (FK) representation of the $q$-state Potts
model~\cite{Fortuin1972}. In this mapping,  the ordering transition of the
Potts model corresponds to a correlated \emph{bond percolation} problem. For $q
> q_c(d)$ (with $q_c=2$ for $d\geq 4$), the ordering transition of the Potts
model is of first order, and hence so is the concomitant FK bond percolation
transition. Another type of percolation models with a known first order
transition are so-called $k$-core and closely related rigidity percolation
problems~\cite{Moukarzel1997,Duxbury1999,Schwarz2006}. Here, despite local
update rules, the percolation phenomenon itself cannot be detected without
``postprocessing'', which both requires an extensive number of checks and
complicates a reversible dynamical process interpretation.

 Pauli percolation can be easily generalized to a
non-equilibrium growth process, e.g. by simply removing the detailed balance
implied by the configuration weights, and retaining only the relative rates for
particle addition. In general, there is a huge family of non-equilibrium
prescriptions which ``generalize'' a given equilibrium distribution. The
equilibrium process -- besides widening the purview of applications from the
exclusively non-equilibrium cases -- leads to a great simplification in the
analysis. It can be efficiently studied numerically on a wide range of graphs
and lattices, and therefore incorporates geometric structures and
inhomogeneities which may be called for in real-life applications. On
sufficiently regular graphs, it can be studied \emph{exactly} with standard
analytical methods. This in particular obviates worries about crossovers on
absurdly long lengthscales or anomalously small critical exponents
\cite{daCosta2010,Grassberger2011}. Approximations such as geometry-free
prescriptions for product-rule percolation also become unnecessary.

In summary, Pauli percolation is a simple, physical, natural, transparent and
tractable novel percolation problem exhibiting an intriguing phenomenology. It
holds great promise as a benchmark problem across the range of disciplines
interested in percolation problems, ranging from condensed matter via
biological systems and real-world networks to epidemic disease outbreaks.
\acknowledgments{We are greatful to R.~D'Souza, P.~Grassberger, M.~Hastings and
L.~Chayes for helpful discussions. We would also like to acknowledge our
collaboration with O.~Derzhko, J.~Richter and A.~Honecker on closely related
work. This work was supported in part by the NSF through grant DMR-0748925
(K.~S.). }

\putbib
\end{bibunit}

\setcounter{equation}{0} \setcounter{figure}{0}
%
\makeatletter 
\def\tagform@#1{\maketag@@@{(S\ignorespaces#1\unskip\@@italiccorr)}}
\makeatother
\makeatletter \makeatletter \renewcommand{\fnum@figure}
{\figurename~S\thefigure} \makeatother

\renewcommand{\bibnumfmt}[1]{[S#1]}
\renewcommand{\citenumfont}[1]{S#1}


\newpage
\begin{bibunit}[Science]

\section{Supplementary Material}
\label{sec:supp}


\subsection*{S1. Exact solution for Pauli percolation on a Cayley tree}

We will use the following definitions: a level-$k$ Cayley tree of
coordination number $z$  is constructed recursively by connecting a root site
to ${z-1}$ identical level-${(k-1)}$ trees -- until level $0$ is reached. We
will refer to level-$0$ sites as leaves; they constitute the outer boundary
of the Cayley tree. The so-called \emph{wired} boundary conditions which we
will consider here are equivalent to establishing additional connections
between all boundary sites~\cite{Chayes1986b,Chayes1999a}. On the other hand,
the free boundary conditions correspond to the leaves having only a single
neighbor, the one at the next level.

We write the partition function of the two-color percolation problem for a
level-$k$ Cayley tree (here we present the case of $z=3$) as a sum of
contributions corresponding to the `fate' of its root site:
\begin{equation}
Z_{k}=E_{k}+F^\text{u}_{k}+F^\text{i}_{k}+U_{k}+I_{k},
\end{equation}
where $E_{k}$, $F^\text{u/i}_{k}$, $U_{k}$  and $I_{k}$ account for all
configurations in which the root site at level $k$ is, respectively, empty or
belongs to a finite uninfected/infected, giant uninfected ($U$) or giant
infected ($I$) cluster. We call a cluster \emph{infected} if it contains a
single red site; a cluster is referred to as \emph{giant} if it contains both
the root and a boundary site, or as \emph{finite} otherwise. By attaching two
level-$k$ trees to a new root site at level $k+1$ and denoting
$H_{\text{k}}=E_{k}+F^\text{u}_{k}$ we arrive at the following recursion
relations
\begin{equation}
\begin{split}
E_{k+1}&=(1-2\tilde{p})Z^{2}_{k},\\
F^\text{u}_{k+1}&=\tilde{p} H^{2}_{k},  \qquad
F^\text{i}_{k+1}=\tilde{p}(H^{2}_{k}+2F^\text{i}_{k}H_{k}),\\
U_{k+1}&=\tilde{p} \left[2U_{k}H_{k}+U_{k}^{2}\right],\\
I_{k+1}&=\tilde{p}\left[2I_{k}H_{k}+2U_{k} H_{k}
+ 2U_{k} F^\text{i}_{k}
+U_{k}^2+I_{k}^2\right],
\end{split}
\label{eq:Cayley_recursion}
\end{equation}
where $\tilde{p}$ is a priori probability of site being occupied and colored
red or green as follows from the main text. Note that the term containing
$I_{k}^2$ in the last line implies, somewhat counterintuitively, that two giant
infected clusters can be merged. In fact, this is a consequence of the 'wired'
boundary conditions: these are two parts of the \emph{same} cluster which are
already connected via boundary sites. Essentially, wired boundaries imply that
there may only exist a single giant cluster.  For the same reason, no ${U_{k}}
{I_{k}}$ terms are possible. (Note that this situation is reversed for free
boundary conditions.) The partition function of a $(k+1)$-level tree is
\begin{equation}
 Z_{k+1}=(1-2\tilde{p})Z^{2}_{k}
 +2\tilde{p}Z_{k}\left(H_{k}
 + U_{k}\right)-2\tilde{p}I_{k}U_{k} +\tilde{p}I_{k}^{2}
\end{equation}
We define $P^\text{u}_{\infty}=\displaystyle
\lim_{k\to\infty}{U_{k}}/{Z_{k}}$ and $P^\text{i}_{\infty}=\displaystyle
\lim_{k\to\infty}{I_{k}}/{Z_{k}}$ to be probabilities that the root site of a
large tree is connected to its boundary via an uninfected and infected
clusters respectively. If $\tilde{p}<4/9$, the only real solution of the
resulting equations is $P^\text{u}_{\infty}=P^\text{i}_{\infty}=0$ -- in
other words, there is no percolation. If $\tilde{p}\geq4/9$, however, two
additional solutions emerge:
\begin{equation}
P^{\pm}_{\infty}=P^{\text{i},\pm}_{\infty}=\frac{1}{2}\pm\frac{3}{2}\sqrt{1-\frac{4}{9\tilde{p}}},
\qquad
P^\text{u}_{\infty}=0.
\label{prob_infinite2}
\end{equation}
As has been pointed in the main text, the fact that $P^\text{u}_{\infty}$
remains zero even after the onset of percolation is rather obvious since the
probability of a large cluster to remain uninfected tends to zero with its
size.

In the same manner we can compute other quantities such as the probability of
a given site being empty $P_\text{e}=\lim_{k\rightarrow\infty} E_{k}/Z_{k}$,
as well well as being occupied and belonging to either an uninfected or
infected finite cluster,
$P^{\text{u},\text{i}}_\text{f}=\lim_{k\rightarrow\infty}
F^{\text{u},\text{i}}_{k}/Z_{k}$. If there is no percolation
($P^{\text{u}}_{\infty}=P^{\text{i}}_{\infty}=0$), these expressions are
given by
\begin{eqnarray}
P_\text{e}&=&\frac{\sqrt{1-4 \tilde{p}^2}}{2 \tilde{p}+1} \nonumber \\
P^{\text{u}}_\text{f}&=&\frac{4 \tilde{p}^2+\sqrt{1-4 \tilde{p}^2}-1}{2 \tilde{p}
(2 \tilde{p}+1)}\nonumber\\
P^{\text{i}}_\text{f}&=&\frac{1}{2\tilde{p}} \left(1-\sqrt{1-4 \tilde{p}^2}\right).
\label{eq:cayley_prob1}
\end{eqnarray}
Above the percolation threshold,  $P^{\textbf{i}}_{\infty}\neq0$, the
expressions become rather cumbersome:
\begin{eqnarray}
P_\text{e}&=&\frac{\sqrt{2} (1-2 \tilde{p})}{\theta_{\pm}(\tilde{p})}\nonumber\\
P^{\text{u}}_\text{f}&=&\frac{1}{4 \tilde{p} (14 \tilde{p}-9)}\left[
9 \tilde{p}-14 \tilde{p}^2
+ 3  \sqrt{2} \left(\tilde{p}-1\right)\theta_{\pm}(\tilde{p}) \right.
\nonumber\\
&&\left. \pm\sqrt{\tilde{p}}\sqrt{9
\tilde{p}-4} \left(9-\sqrt{2} \theta_{\pm}(\tilde{p})- 14\tilde{p}\right)
\right]\nonumber\\
P^{\text{i}}_\text{f}&=&\frac{3 \tilde{p}\mp\sqrt{\tilde{p}\left(9 \tilde{p}-4\right)}
-\sqrt{2} \theta_{\pm}(\tilde{p})}{4 \tilde{p}}
\label{eq:cayley_prob2}
\end{eqnarray}
with
$$
\theta_{\pm}(\tilde{p})=\sqrt{\tilde{p} \left(6-11 \tilde{p}
\pm\sqrt{\tilde{p}\left(9 \tilde{p}-4\right)} \right)}.
$$
These solutions are used in the derivation of the corresponding
probabilities for the $z$-regular random graph.

\subsection*{S2. Site/Edge addition to a $z$-regular random graph}

Using the cavity method we can now obtain the results for the full-space
$z$-regular random graph. In the cavity method the addition of a bulk site or
edge is equivalent to connecting $z$ or $z-1$ roots of independent level-$k$
Cayley trees (see Figure~\ref{fig:bethe}) in the main text). In other words the
bulk site or edge of $z$-regular random graph is equivalent to the central site
or edge of a uniform Bethe lattice. The quantities for the site-centered case,
analogous to those given by Eqs.~(S\ref{eq:Cayley_recursion}) for a rooted
tree, can be written as
\begin{eqnarray}
{\mathcal{E}}&=&\left(1-2\tilde{p}\right)Z_{{k}}^{3},
\quad
{\mathcal{F}}^\text{u}= \tilde{p} H_{k}^{3},\quad
{\mathcal{F}}^\text{i} = \tilde{p}\left(H_{k}^{3}
+ 3F^\text{i}_{k}H_{k}^{2}\right),\nonumber\\
\mathcal{I}&=& 3\tilde{p}I_{k}H_{k}^{2}+
3\tilde{p}I_{k}^{2} H_{k}
+ \tilde{p}I_{k}^{3}
\end{eqnarray}
where we have discounted all configurations where the giant cluster is
uninfected -- we have already seen that they have vanishing relative
contribution. The partition function is then
\begin{multline}
\mathcal{Z}={\mathcal{E}}+{\mathcal{F}}^\text{u}+{\mathcal{F}}^\text{i}+{\mathcal{I}}\\
 =(1-2\tilde{p})Z^{3}_{k}
 + 3\tilde{p}Z_{k}H_{k}^2
 -\tilde{p}H_{k}^3 +3 \tilde{p}I_{k}^2 H_{k} + \tilde{p}I_{k}^{3}
\label{eq:partition_full}
\end{multline}

Another way of constructing a uniform lattice is by adding an edge between
two root sites of Cayley trees. The corresponding quantities in this case
become:
\begin{equation}
{\tilde{\mathcal{H}}} = H_{{k}}^{2},
\quad
\tilde{\mathcal{F}}^\text{i}=2F^\text{i}_{k}H_{k},
\quad
\tilde{\mathcal{I}}= 2I_{k}H_{k}+I_{k}^{2}.
\end{equation}
The meaning of these quantities in the case of edge addition is as follows:
$\tilde{\mathcal{H}}$ is the number of all configurations where each of the
(former root) sites connected by the new edge was either empty or belonged to
a finite uninfected cluster; $\tilde{\mathcal{F}}^\text{i}$ counts all
configurations where one of these sites belonged to a finite infected cluster
while the other was either empty or a part of a finite uninfected cluster;
$\tilde{\mathcal{I}}$ counts configurations where either one or both sites
belonged to a giant (infected) cluster. Once again, we discount the
configurations where the giant cluster is uninfected. The corresponding
partition function is
\begin{equation}
\tilde{\mathcal{Z}} = \tilde{\mathcal{H}}+
\tilde{\mathcal{F}}^\text{i}+ \tilde{\mathcal{I}} =
2 Z_k H_k - H_k^2 + I_k^2.
\label{eq:bond_partition}
\end{equation}

Using these quantities, we can calculate various probabilities in the same
fashion as in the previous section for the root site -- see
Eqs.~(S\ref{eq:cayley_prob1},S\ref{eq:cayley_prob2}). Note that the value of
parameter $\tilde{p}=4/9$ at which the percolation solution
$\mathcal{P}_{\infty}\neq0$ first emerges is not affected by such calculation,
albeit the value of the percolation probability itself
 changes: $\mathcal{P}_{\infty}\neq {P}_{\infty}$.

The importance of merging the rooted Cayley trees into a Bethe lattice in
these two different ways will become clear in the next section dedicated to
calculating the free energy. This will allow us to circumvent the inherent
problem of evaluating extensive thermodynamic potentials on the Bethe
lattice, where the number of boundary sites is a finite fraction of the total
system.

\subsection*{S3. Bulk free energy}

In contrast with continuous phase transitions, first order transitions do not
occur when the non-trivial solution for the order parameter first appears as
this normally signifies only the emergence of a \emph{metastable} state.
Therefore, in order to determine the actual transition point in this case,
one needs to study the free energy; the transition occurs when the free
energy associated with an ordered state becomes smaller than that for the
disordered state. This seemingly straightforward test becomes problematic on
a Bethe lattice due to the aforementioned issue of an extensive size of the
boundary. While this problem had been widely discussed in the literature
-- see
e.g.~\cite{Mueller-Hartmann1974,Eggarter1974,Wang1976,Peruggi1983,Gujrati1995,Duxbury1999}
--  none of the recipes proposed there are applicable (or even meaningful)
for the case of percolation. Specifically, our percolation model is a
counting problem and does not have any sensible notion of energy associated
with it, hence no derivatives with respect to external fields can be used to
define any thermodynamic potentials here, unlike, e.g. in the context of the
Potts model~\cite{Peruggi1983}. One could define the limit of the free energy
per \emph{internal} site following the approach of
Ref.~\onlinecite{Gujrati1995}, yet this quantity is not helpful either: not
only is such free energy \emph{always} minimized in the percolating phase, it
is not even continuous across the (putative) transition into this phase.
Since the actual free energy must be continuous across any phase transitions,
it is clear that the aforementioned free energy per internal site is not the
right quantity to look at in our case. (Naturally, the total free energy
defined via the logarithm of the partition function is a continuous function
of its parameters but contains  an extensive boundary contribution.) In
short, the failure of this approach signifies a simple fact that the free
energy cannot be associated solely with an internal site or an internal bond
of a Bethe lattice, and the presence of an extensive boundary prevents one
from meaningfully distributing its `shares' between them. Specifically, a
choice of boundary conditions (e.g. free vs. wired) dramatically changes the
ratio between the number of bonds and the number of sites in the system. Note
that this issue does not arise in the context of continuous phase transitions
since those always coincide with the emergence of a non-trivial solution for
the order parameter.

The problem with a meaningful definition of the free energy is cured by considering
a $z$-regular random graph instead of a Bethe lattice. The two are locally
equivalent to one another, yet the random graph lacks a distinct boundary. This
in turn fixes the bond to site ratio of in the system to $z/2$. We can then use
the prescription outlined in Ref.~\cite{Barre2007} to write the free energy per
added site as ${\mathcal{F}  = \lim_{k\to\infty}\left[
-\ln{{\mathcal{Z}}/Z_k^3}+z/2\ln{{\tilde{\mathcal{Z}}}/Z_k^2}\right]}$ where
the first term corresponds to the the free energy of an internal site of a
Bethe lattice defined similarly to Ref.~\cite{Gujrati1995} while the second
term corrects for the fact that adding a site to a $z$-regular graph creates
$z$ new edges, and hence $z/2$ existing edges should be removed to maintain the
graph's regularity.

Using expressions for $\mathcal{Z}$, $\tilde{\mathcal{Z}}$ and $Z_{k}$ given
by Eqs.~(S\ref{eq:partition_full}) and (S\ref{eq:bond_partition}) of the
previous section, we find the expressions for free energy corresponding to
all three solutions for $\mathcal{P}_\infty$ on a $3$-regular random graph:
\begin{subequations}
\begin{equation}
\mathcal{F}_{0} = -\frac{1}{2}\ln{2} -\ln{\tilde{p}}
+ \frac{1}{2}\ln{\left(\frac{1}{\sqrt{1-4\tilde{p}^2}}-1\right)}
\label{eq:free_energy_nonperc}
\end{equation}
\begin{multline}
{\mathcal{F}}_{\pm}=-\frac{3}{2}\ln2 - \ln\left[26\mp 6 \sqrt{\tilde{p}(9 \tilde{p}-4)}
-46 \tilde{p}\right.
\\
\left.
+\sqrt{2} \left(11 -6/\tilde{p}\right) \theta_{\pm}(\tilde{p})
 \pm\sqrt{9 -4/\tilde{p}} \left(6-\sqrt{2} \theta_{\pm}(\tilde{p})\right)\right]
 \\
 +\frac{3}{2} \ln \left[22 +
 \frac{5 \sqrt{2} \theta_{\pm}(\tilde{p}) -12}{\tilde{p}}
 \right.
 \\
 \left.
 \pm \sqrt{9 -4/{\tilde{p}}}\left(\sqrt{2}\theta_{\pm}(\tilde{p})/\tilde{p}
   - 2 \right)\right].
\label{eq:free_energy_perc}
\end{multline}
\label{eqs:free_energy}
\end{subequations}
 These expressions are plotted in Figure~\ref{fig:percolation} of
the main text.

\subsection*{S4. Average cluster size and cluster size distribution}

\begin{figure}
\centering
\includegraphics[width=.9\columnwidth]{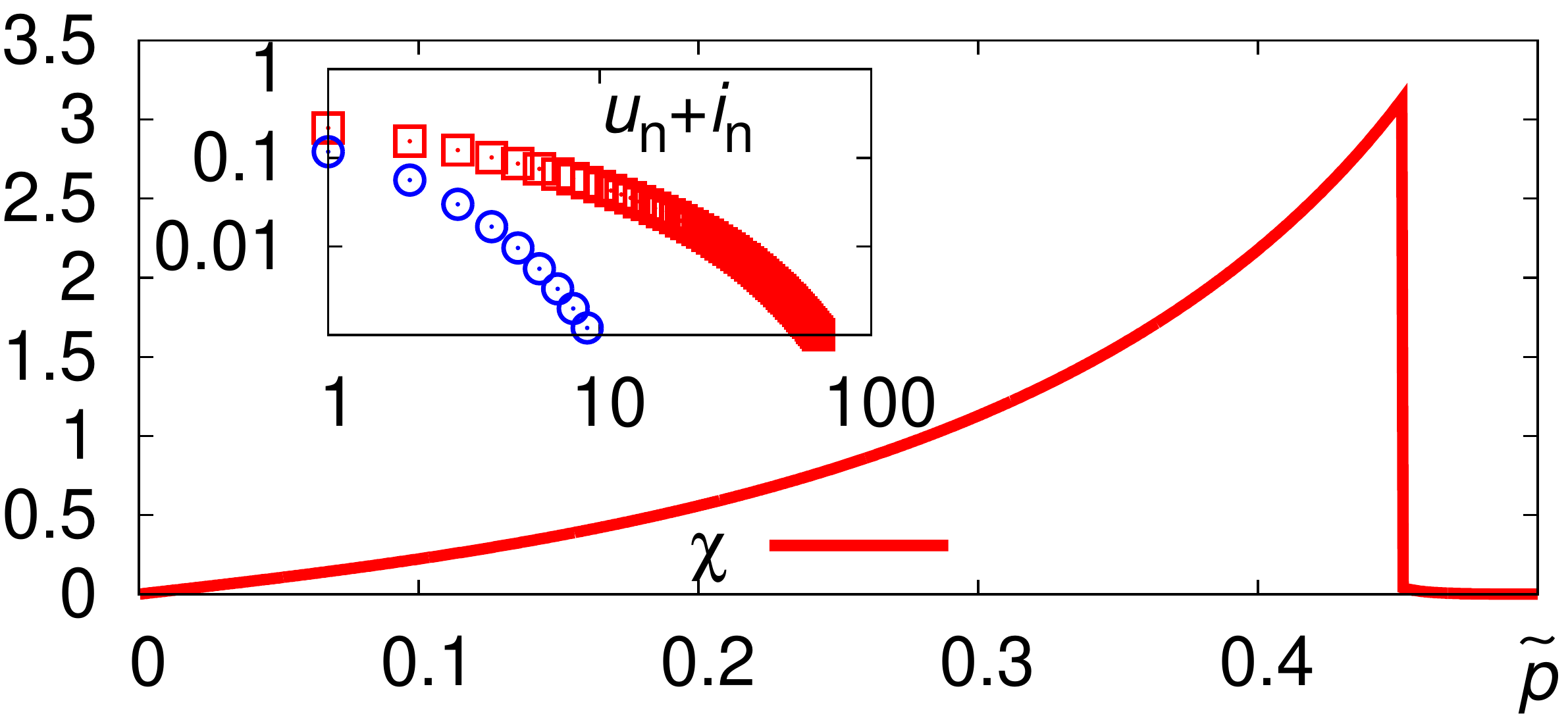}
\caption
{Average size of a finite cluster $\chi(\tilde{p})$.
The inset shows the cluster size distribution $f(n) = u_n + i_n$ below percolation,
at $\tilde{p}=0.45<\tilde{p}_{c}$ (red)
and above percolation, $\tilde{p}= 0.452>\tilde{p}_{c}$ (blue).
}
\label{S4.fig01}
\end{figure}

Having obtained the solutions of the recursion relations for a single rooted
Cayley tree, we have access to other physical quantities of interest such as
the average cluster size or the cluster size distribution in the same recursive
manner. For example, the expected size of a cluster containing the root site
but not any leaves is given by
\begin{equation}
\chi_{{k}}=\frac{f_{k}F_{k}}{Z_{k}}
= \frac{f^\text{i}_{k}F^\text{i}_{k}+f^\text{u}_{k}F^\text{u}_{k}}{Z_{k}},
\end{equation}
where $f_{k}$ is the average cluster size in non-percolating configurations,
$F_{{k}}$ is the weight of corresponding configurations. As before, labels `i'
and `u' indicate infected or uninfected clusters. The $k\to\infty$ limit of
this quantity plays the role of susceptibility in conventional percolation
problems. With minimal effort, it also can be found for the uniform Bethe
lattice; the result is shown in Figure~S1. Since in our case the transition is
first order, this quantity does not diverge at the transition
$\tilde{p}=\tilde{p}_{c}$.

The cluster size distribution for a rooted tree can be obtained from the total
weight of configurations where the site at level $k$ belongs to a finite
cluster
\begin{equation}
F_{k}=F^\text{i}_{k}+F^\text{u}_{k}=\sum_{n}F^\text{i}_{k}(n)
+\sum_{n}F^\text{u}_{k}(n).
\end{equation}
Here $F^{\text{u,i}}_{k}(n)$ are the weights of configurations where the root
site belongs to a cluster of size $n$. The corresponding probabilities for a
site to belong to an uninfected/infected $n$-site cluster are
$(u,i)_{n}=\displaystyle \lim_{k\to\infty} F^\text{u,i}_{k}(n)/{Z_{k}}$ and may
be also found from recursion relations
\begin{subequations}
\begin{equation}
u_{n}=\tilde{p} \eta\left(\sum^{n-2}_{k\geq1}u_{k}u_{n-k-1}+2u_{n-1}P_\text{e}\right),
\end{equation}
\begin{multline}
i_{n}=\tilde{p} \eta\left(\sum^{n-2}_{k=1}u_{k}u_{n-k-1}
+2\sum^{n-2}_{k=1}i_{k}u_{n-k-1}\right.
\\
+2P_\text{e}i_{n-1}+2P_\text{e}u_{n-1}
\Bigg).
\end{multline}
\end{subequations}
Here the probabilities for a site to form an isolated cluster or to remain
unoccupied are, respectively,
\begin{equation}
u_{1}=i_{1}=\tilde{p}\eta P^2_\text{e},\qquad u_{0}=i_{0}=P_\text{e}
\end{equation}
and $\eta = \lim_{k\to\infty} Z_{k}^2/Z_{k+1}$.

The generating functions for both sequences are
\begin{equation}
\mathcal{G}^\text{u}(x)=\sum_{n\geq0}u_{n}x^{n}=\frac{1-\sqrt{1-4x\tilde{p}\eta P_\text{e}
}}{2x\tilde{p}\eta}
\end{equation}
and
\begin{equation}
\mathcal{G}^\text{i}(x)=\sum_{n\geq0}i_{n}x^{n}= \frac{P_\text{e}
+x\tilde{p}\eta\left[{\mathcal{G}^\text{u}}(x)\right]^2}
{1-2x\tilde{p}\eta{\mathcal{G}^\text{u}}(x)},
\end{equation}
and its series expansion yields $u_{n}$ and $i_{n}$.

The cluster size distribution (inset of Figure~S\ref{S4.fig01}) has an
exponential cutoff for large clusters both below and above the percolation
transition. After the appearance of the giant component the cutoff
discontinuously shifts to smaller cluster sizes. A similar cluster size
distribution can in principle be obtained for a uniform Bethe lattice yet in
practice the recursion relations become extremely unwieldy.

\subsection*{S5. Details of numerical simulations}

\begin{figure}
\centering
\includegraphics[trim=1.1cm 11.5cm 1.1cm 1cm, clip=true,angle=0,width=\columnwidth]{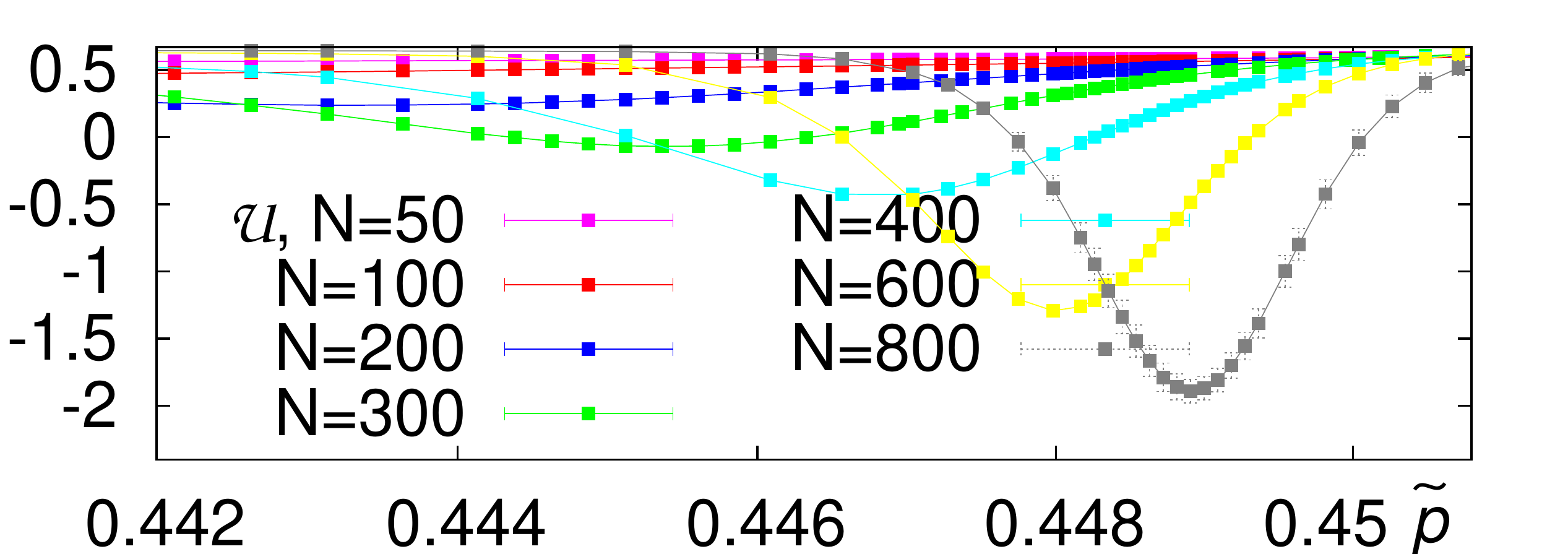}
\includegraphics[trim=1.1cm 11.5cm 1.1cm 1cm, clip=true,angle=0,width=\columnwidth]{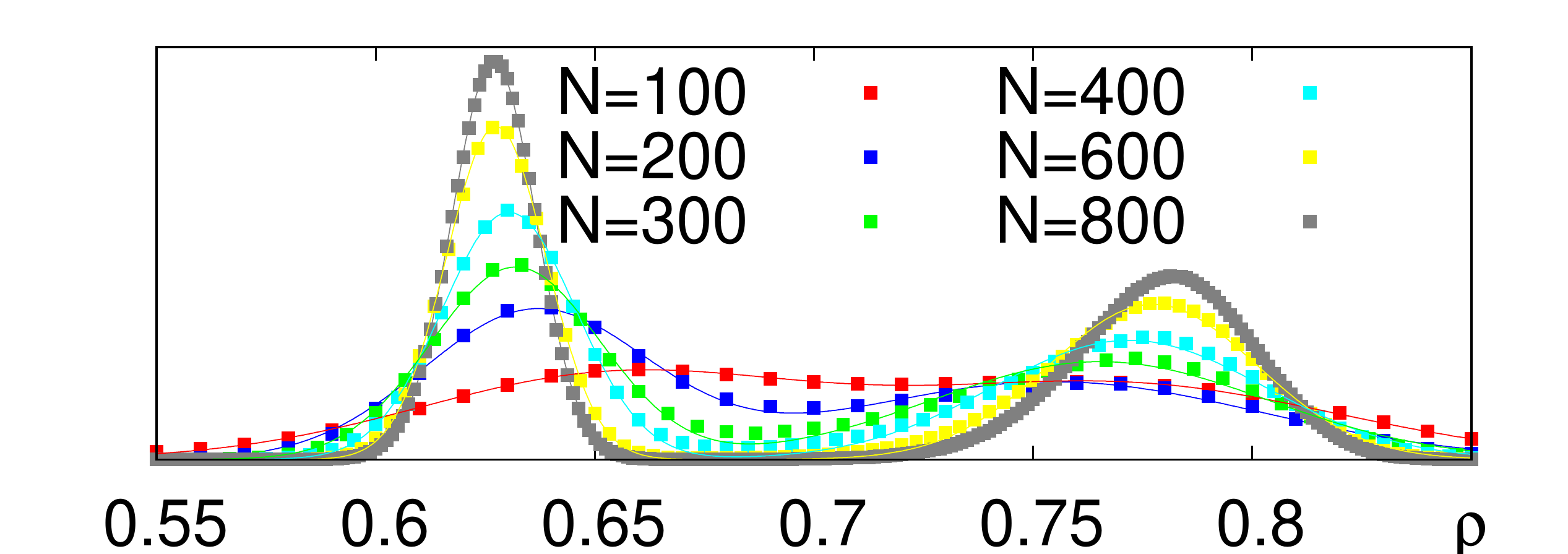}
\caption
{a) Fourth-order Binder cumulant of density. b) Histograms of density at point where peaks
are equally weighed; lines indicate fit to a pair of Gaussian functions.
}
\label{SFig02}
\end{figure}

We use a Metropolis Monte-Carlo algorithm developed in \cite{Maksymenko2012} on
 graphs with up to ${N}=820$ sites. The algorithm works in the
grand-canonical picture where a chemical potential $\mu$ controls the density
of particles $\rho$ and allows it to fluctuate. The statistical weight is
$W=\exp{(\mu n)} \prod(n(\mathcal{C}_{i})+1)$. The chemical potential $\mu$ is
directly related to the a priori probability $\tilde{p}$ via
$\mu=\ln\left[{\tilde{p}/(1-2\tilde{p}})\right]$. At every step we randomly
choose a site (or group of sites) and if it is empty (occupied), propose to
occupy (empty) it. The new configuration is accepted with a Metropolis
probability. We use up to $2\times10^6$ steps for equilibration which are then
followed by $2\times10^6$ steps for every measurement round. The plots
presented in the report are based on averaging over up to $30$ measurements
with a new realization of random graph for every measurement. The expander
nature of the graph and the long range of underlying interactions lead to
strong hysteresis. To reduce hysteresis, we have employed an exchange
Monte-Carlo procedure by simulating the system at different values of
$\tilde{p}$ and allowing exchanges of configurations between them. To control
hysteresis effects, we have performed simulations starting from empty or
occupied lattices as initial conditions. We present results for system sizes
which show no hysteresis. In Figure~S\ref{SFig02} we present a fourth-order
Binder cumulant ${\mathcal{U}}$ defined as.
\begin{equation}
{\mathcal{U}}=1-\frac{\langle \rho^4 \rangle}{3\langle \rho^2 \rangle^2}
\label{eq:Binder}
\end{equation}
Minima of this quantity indicates the location of a phase transition, their
extrapolation to the thermodynamic limit is plotted in the main text.

\putbib
\end{bibunit}
\end{document}